\documentclass[12pt]{article}

\usepackage{amsmath,amssymb}
\usepackage{graphicx}
\usepackage{caption}
\usepackage{color}
\usepackage{dcolumn}
\usepackage{bm}
\usepackage{float}
\usepackage{hyperref} 
\usepackage{ragged2e}
\usepackage{natbib}
\hypersetup{ colorlinks = true, citecolor = black, urlcolor = blue}
\usepackage{xcolor}

\definecolor{background-color}{gray}{0.98}

\title{Change Detection in Dynamic Attributed Networks}

\author{Isuru Udayangani Hewapathirana \thanks{School of Mathematics and Statistics, University of Canterbury, Christchurch, New Zealand.}}

\date{}
\begin{document}
\maketitle

\begin{center}
\subsubsection*{\small Article Type:}
Overview

\hfill \break
\thanks

\subsubsection*{Abstract}

\justify
A network provides { powerful means} of representing complex relationships between entities by abstracting entities as vertices, and relationships as edges connecting vertices in a graph.  Beyond the presence or absence of relationships, a network may contain additional information that can be attributed to the entities and their relationships. Attaching these additional attribute data to the corresponding vertices and edges yields an attributed graph. Moreover, in the majority of real-world applications, such as online social networks, financial networks and transactional networks, relationships between entities evolve over time. 
\justify
Change detection in dynamic attributed networks is an important problem in many areas, such as fraud detection, cyber intrusion detection and health care monitoring. It is a challenging problem because it involves a time sequence of attributed graphs, each of which is usually very large and can contain many attributes attached to the vertices and edges, resulting in a complex, high dimensional mathematical object.
\justify
In this survey we provide an overview of some of the existing change detection methods that utilize attribute information. We categorize these methods based on the levels of structure in the graph that are exploited to detect changes. These levels are vertices, edges, subgraphs, communities and the overall graph. We focus our attention on the strengths and weaknesses of these methods, including performance and scalability. Finally we discuss some publicly available dynamic network datasets and give a brief overview of simulation models to generate synthetic dynamic attributed networks.

\end{center}

\clearpage

\renewcommand{\baselinestretch}{1.5}
\normalsize

\clearpage

\section*{\sffamily \Large GRAPHICAL TABLE OF CONTENTS} 

\begin{figure}[H]
	\centering
	\includegraphics[trim = 0mm 0mm 0mm 0mm, scale=0.5]{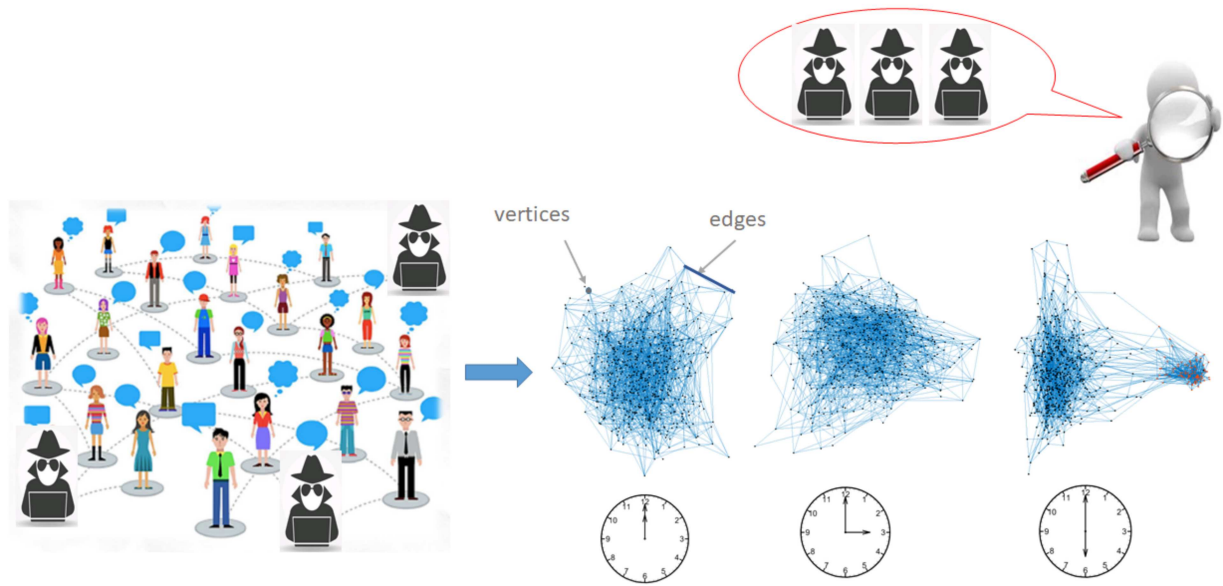}
	\caption{{Change Detection in Dynamic Attributed Networks}}
	\label{graphicalAbstract}
\end{figure}

\section{\sffamily \Large INTRODUCTION} 
\label{Introduction}

Rapid developments in the field of information technology have resulted in a proliferation of datasets over the past few decades. Examples include supermarket transaction data, credit card records, medical records, and records of telephone call information. The availability of large datasets led to a high demand for new methods to extract useful information. Thus the field of \textit{data mining} started to develop. Data mining involves methods for exploring, analysing, and modelling large datasets to find valid, novel, and interpretable patterns \citep{fayyad1996data}. The area of data mining which attempts to discover unusual occurrences in dynamic (time-varying) datasets is called \textit{change detection}. Detecting changes provides early warnings of events that can have serious impact. For example, early detection of an abrupt change in the stock market could help improve
decision making and avoid possible losses for businesses due to wrong decisions
and weak trading systems. The majority of traditional data mining methods assume entities in a dataset to be independent, and ignore the natural relationships among them \citep{huang2014clustering}. However, entities in most real-world datasets are not independent, but are related to each other in different ways. For accurate change detection, it is vital to consider these underlying relationships. For example, the extent to which a transaction is fraudulent in credit card data can be identified more accurately by considering the usual purchasing behaviour of the card-holder as well as the usual purchasing behaviours of other card holders who bought the same product \citep{singh2015fraud}. The need for regarding such inherent dependencies among entities has aroused interest for network-based change detection methods. 

\begin{figure}[h]
	\centering
	\includegraphics[trim = 0mm 0mm 0mm 0mm, scale=1.4]{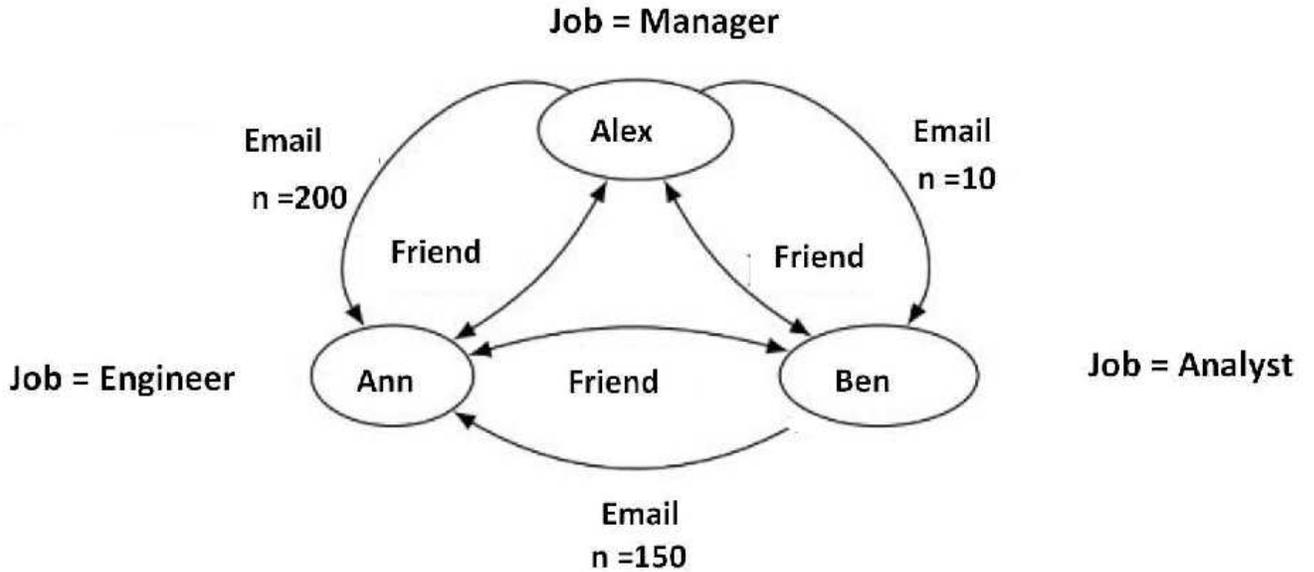}
	\caption{{Illustration of an attributed network.}}
	\label{illustrateNetwork}
\end{figure}

A network is a collection of entities, that have inherent relationships. Some examples include a social network of friendships among people, a communication network of company employees connected by phone calls, emails or text messages, and a biological network of neurons connected by their synapses.  A network can be mathematically conceptualized as a graph by associating entities with vertices, and relationships with edges connecting vertices in the graph. For example, in the graph representation of a social network like Facebook, vertices may represent friends and edges represent friendship connections. It is also possible to capture more information by attaching additional attribute data to the vertices and edges, resulting in an attributed graph. An illustrative example of a small attributed graph of three vertices is shown in Figure \ref{illustrateNetwork}. Each vertex is tagged with attribute information about the job role of the corresponding employee, while the edges are tagged with attribute information based on the type of relationship (whether the relationship is based on friendship or email sending patterns). An edge can be further weighted to quantify the corresponding relationship. In the example given in Figure \ref{illustrateNetwork}, the edges representing email sending patterns are weighted to represent the number of emails.  

Most real-world networks evolve as time progresses. That is, the entities and their relationships keep evolving with time. This type of relational data can be represented as a dynamic network. For example, a communication network of a company is a dynamic network because new employees (entities) join the network and communication patterns (relationships) are modified continuously. A dynamic network can be represented as a time sequence of graphs, each representing the entities (as vertices) and their relationships (as edges) at a given time instant. 

\begin{figure}[h]
	\centering
	\includegraphics[trim = 0mm 0mm 0mm 0mm, scale=1.5]{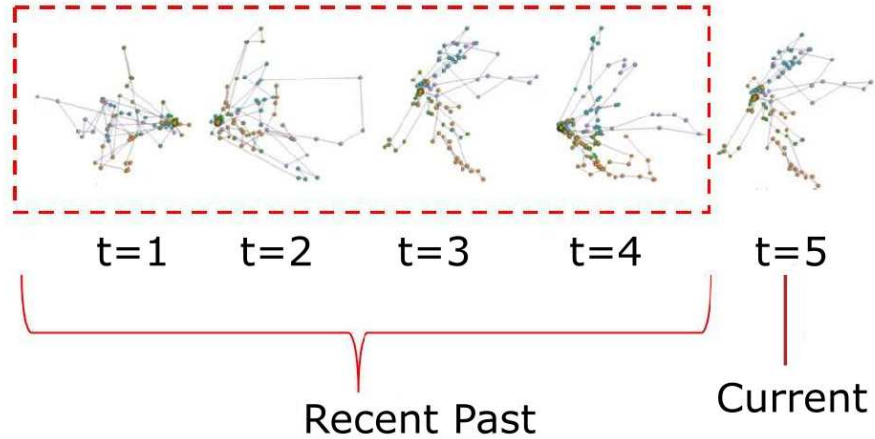}
	\caption{{\textbf{Illustrative example of change detection}. Each graph represents the behaviour of the entities and their relationships at the corresponding time instant. A sequence of graphs inside the window containing the recent past time instants characterises profile behaviour of entities and their relationships. A change detection method compares the current behaviour of the overall graph, vertices, edges, or subgraphs with their profile behaviour. \label{illustartChangeDetection}}}
\end{figure}

Change detection in dynamic networks is the process of continuously monitoring a network for deviations in the behaviour of entities and their relationship structure. Figure \ref{illustartChangeDetection} illustrates this based on a toy example. The graph obtained at the current time instant represents the {current} behaviour of entities and their relationships, while the sequence of graphs inside the window containing the \textit{recent past} time instants, characterise \textit{profile} behaviour. A general change detection method aims to build a model based on the profile behaviour of the graph during the recent past, and search for deviations in the behaviour of the current graph with respect to the model. The majority of change detection methods monitor a \textit{representative summary} extracted from the overall graph, or parts of the graph, such as vertices, edges or subgraphs\footnote{ A subgraph is a subset of vertices and edges in the graph.} to characterise behaviour and analyse deviations. For example, in order to detect fraudulent collaborations in social networks, \cite{yu2014glad} track the properties of communities, while \cite{neil2013towards} observe the appearance of unlikely edges to detect intrusions in a computer network. The choice of {unit} (either overall graph, vertices, edges or subgraphs) to track over time to detect changes, mainly depends on the given application.  

Change detection in dynamic networks has many vital applications:
\begin{itemize}
	\item Social network analysis:\\
	Changes in popular social networks such as Facebook, Twitter, LinkedIn or Youtube may arise as a result of individuals or groups of individuals making deviations in their usual relationship patterns. For example, a group of fraudsters in an on-line auction network who previously had no relationships may suddenly collaborate to boost the reputation of their product. Such changes will be reflected as changes in the corresponding dynamic network, so change detection enables detection of those fraudsters \citep{pandit2007netprobe}.
	\item Financial market analysis:\\
	A financial network represents the dynamic inter-relationships between the world's various financial markets. Changes in international trade or countries' stock markets at a certain point in time, correspond to changes in the structure of the corresponding financial network \citep{durante2014bayesian} . Such changes should be detected early to minimize financial crisis situations.  
	\item Network traffic monitoring:\\
	A dynamic computer network can be represented as a time sequence of edge attributed graphs, where the vertices represent the computers, and the edges represent the volume of traffic between each pair of computers. A change in the usual amount of traffic between a pair of computers may indicate a fault in the system or even an intrusion. Such changes can be detected by continuously tracking the edges over time \citep{sun2006beyond}. 
	\item Analysis of earth-science data:\\
	Earth scientists detect natural disasters by analysing data about variables such as vegetation cover, temperature, and pressure at certain locations over time. Natural disasters such as wildfires correspond to abrupt changes in the time-series of these variables. However, these variables are most often inherently correlated with each other. Accounting for these inherent correlations, by formulating a network, and applying network-based change detection methods, increase the detection ability of such natural disasters \citep{cheng2008robust}. 
	\item Connectomic applications:\\
	Connectomics involves the study of inter-relationships among different regions of the brain. Inter-relationships between brain regions can be observed over time, and represented as a dynamic network. A sudden neurological failure such as an epilepsy attack, will correspond to a change in the structure of the network {\citep{durante2014nonparametric}}. Hence, change detection can be used to improve diagnosis and treatment of such medical conditions.
\end{itemize}

In order to accurately detect changes, we need to consider the following issues:

{\textit{Scale and dynamics}}: Many real-world networks are considerably large in size. For example, the Facebook contains more than one billion users,\footnote{\url{http://newsroom.fb.com/company-info/}} and the World Wide Web consists of more than 40 billion web pages\footnote{\url{http://www.worldwidewebsize.com/}}. A graph representing the entity-relationships in these datasets is also large, with billions of vertices and edges. If such data are collected in real time, the speed at which they arrive is also an issue. We require fast methods to study the structure of such graphs at each time instant.

{\textit{Complexity}}: As mentioned previously, the vertices and edges of a graph can be tagged with attributes capturing specific information about entities and their relationships. A change detection method should effectively combine all available and useful information in order to obtain accurate results. However, the inclusion of edge and vertex attributes increases the complexity.

\textit{Lack of prior information}: Data we receive often arrives with no a priori knowledge about the underlying situation. Hence, we have no knowledge about the recent past behaviour, and this should be obtained from the data using a method that is both accurate and computationally efficient \citep{nickel2013tensor}.

In this review we discuss several literature to date on change detection methods in dynamic attributed networks. 

Our paper is organized as follows. First in Section \ref{Other Related Surveys}, we review previous surveys that address the problem of change detection in dynamic networks. In Section \ref{Notation and Terminology}, we define the graph theoretic notations that we utilize throughout the paper. In Section \ref{Categorization of Methods} we first describe the structure of our paper and how we categorize methods for change detection. We further clarify what we consider as attribute information when reviewing past literature.
In the next five subsections, we provide a thorough review of different change detection methods based on the categorization introduced in Section \ref{Categorization of Methods}. In Section \ref{Changes in Global Structure}, we review methods that detect changes in the overall graph. These methods detect points in time where the majority of vertices, edges, subgraphs or communities change their behaviour compared to recent past. In Section \ref{Entity Based Methods}, we review methods that focus on detecting changes in vertex behaviour. Section \ref{Edge Based Methods} discusses methods that detect abnormally evolving edges in the network. A subgraph is a collection of vertices and edges. Section \ref{Sub-graph Based Methods} reviews methods that detect changes in subgraph behaviour. In this category, we separately discuss methods on communities (Section \ref{Community based methods}), which are a special type of subgraph. In Section \ref{Discussion}, we critically assess the strengths and weaknesses of different approaches with respect to computational efficiency and applicability in different situations. We also mention procedures  that can be utilized for evaluating a novel method developed for a dynamic network. Finally in Section \ref{Conclusion}, we provide a concluding summary of our review. 

{
\section{\sffamily \Large NOTATION AND TERMINOLOGY}}
\label{Notation and Terminology}

A network can be mathematically represented as a graph, $G = (V,E{  \subseteq V \times V})$, where $V$ is a set of $n$ \textit{vertices} and $E$ is a set of \textit{edges}. { An edge connects two vertices and can be directed or undirected.} For example, in the case of a simple graph with $n$ vertices, $E$ will contain a list of those edges that are present in the graph. From this edge list, the presence or absence of an edge between each pair of vertices can be encoded as an $n \times n$ binary adjacency matrix, $A$, where each element, $A_{i,j}$, is $1$ if the edge between vertices $i$ and $j$ appears in $E$, and $0$ otherwise. In { an attributed} graph, additional information is assigned to the edges or vertices or both, this information is called a \textit{weight}. A{n attributed} graph can be encoded as an $n \times n$ weighted adjacency matrix, $W$, and $W_{i,j}$ is the non-negative weight of the edge between vertices $i$ and $j$, and $W_{i,j} =0$ means that vertices $i$ and $j$ are not connected by an edge. In this review, we consider graphs where the edge weights are proportional to the strength of the relationship between the corresponding entities. 

A dynamic { attributed} network can be mathematically represented as a time sequence of { attributed} graphs, $G^1,G^2,\ldots,$ where $G^t=(V^t,E^t)$. Since our focus is on a dynamic { attributed} network with a fixed set of entities, we will have $G^t=(V,E^t)$ as a special case.

{
\section{\sffamily \Large OTHER RELATED SURVEYS}}
\label{Other Related Surveys}

\cite{akoglu2015graph} surveys a vast variety of methods that do change detection in dynamic unattributed networks. They categorize these methods based on the type of graph summary extracted as well as the type of change detected. They review feature-based methods, decomposition-based methods, community-based methods and window-based methods. Feature-based methods calculate graph summaries based on graph properties such as the degree distribution of the network. Decomposition-based methods apply matrix or tensor decomposition techniques on the adjacency matrix or tensor, representing the network. Community-based methods monitor the communities in the network to detect structural changes. Window-based methods incorporate the idea of a moving window over the time sequence of graphs to detect changes from the recent past behaviour. They present a thorough treatment on the subject of change detection methods derived for unattributed dynamic networks. However, \citeauthor{akoglu2015graph} do not address the problem of detecting changes in dynamically changing attributed networks.

\cite{aggarwal2014evolutionary} is another survey paper that covers some change detection methods on dynamic networks. The main goal of their paper is to give a comprehensive review on methods for studying the evolutionary behaviour of a dynamic network. They review a few methods that quantify the changes occurring in the vertices and edges. The discussed changes focus on centrality of vertices, behaviour of communities and shortest path distances between vertices. They further discuss the applicability of each method to slowly evolving networks and streaming networks. In addition to change detection methods, \citeauthor{aggarwal2014evolutionary} provide a comprehensive discussion on dynamic network generation mechanisms.

Given a time sequence of graphs, \cite{ranshous2015anomaly} provide a thorough review on methods that detect vertices, edges, subgraphs or time instants that depict irregular evolutionary behaviour compared to the rest of the network.  Their review is organized based on the underlying design principles of those methods, such as those based on community detection, graph compression, matrix decomposition and distance metric calculation. Although \cite{ranshous2015anomaly} cover methods that detect changes in a dynamic network, the main goal of their survey paper is not reviewing change detection methods that exploit attribute information. 

While all the above mentioned survey papers cover different aspects of previous change detection methods, their overall focus is different from ours. In our paper we give a deeper discussion on change detection methods that specifically focus on utilizing attribute information. Furthermore, previous survey papers categorize methods based on the underlying methodologies behind the algorithms employed for change detection. The framework we follow is different. We take into account the levels of structure selected from each graph, to monitor over time for change detection. 

\section{\sffamily \Large CATEGORIZATION OF METHODS}
\label{Categorization of Methods}

As detailed in Section \ref{Introduction}, changes in a dynamic network are deviations from usual activities, and change detection is the process of monitoring a time sequence of graphs to detect such deviations. Most change detection methods follow a common procedure as follows:
\begin{enumerate}
	\item Given a time sequence of graphs, determine the smallest unit in the graph (e.g., a vertex, an edge, a subgraph or the entire graph) that needs to be monitored over time to detect the change. 
	
	\item At each time instant, extract a good summary measure from the graph that accurately represents the behaviour of the smallest unit selected.
	
	\item Calculate the dissimilarity between the summary measure obtained for the current time instant and a summary measure representing the profile behaviour of the unit during the recent past time instants. This dissimilarity measure is the \textit{change score} for the current time instant. 
\end{enumerate}

Based on the unit selected, we divide the change detection methods reviewed into four main categories. They are methods that detect,
\begin{enumerate}
	\item changes in the overall graph (Section \ref{Changes in Global Structure}),
	\item changes in vertices (Section \ref{Entity Based Methods}),
	\item changes in edges (Section \ref{Edge Based Methods}), and
	\item changes in subgraphs (Section \ref{Sub-graph Based Methods}). In this category, we separately discuss methods on communities (Section \ref{Community based methods}), which are a special type of subgraph.
\end{enumerate}
Within each category we { discuss} methods that use vertex attributes and edge attributes. { The attributes that we consider can be additional information about the entities and their relationships in the corresponding network or can be vertex or edge properties calculated from the adjacency matrix of the graph. Examples include quantitative attributes such as vertex degree \citep{borges2011anomaly} and edge weight \citep{ide2004eigenspace}, and categorical attributes such as vertex type \citep{jiang2015link} and edge type \citep{koutra2012tensorsplat}.}  

\subsection{\sffamily \large CHANGES IN THE OVERALL GRAPH}
\label{Changes in Global Structure}

A change in the overall graph occurs when the majority of entities in a network change their usual relationship patterns. For example, \cite{akoglu2010event} observe a mobile communication network over a six months period and discover that most people change their usual communication patterns during the {Christian New Year}. Changes in the behaviour of most vertices or edges in the graph are considered to be a change in the behaviour of the overall graph. Detecting changes in the overall graph is sometimes referred to as \textit{event detection} \citep{ranshous2015anomaly}. 

In a computer network, we can obtain information regarding a set of computers and their interactions over time. Each interaction between a pair of computers contains attribute information such as connection type or number of connections. \cite{ide2004eigenspace} represent such a computer network as a{n attributed} and undirected graph, where the edge { attributes} are the number of connections between two computers at any given time instant. Each { attributed} graph is then represented as a weighted adjacency matrix, $W$. The goal of \cite{ide2004eigenspace}'s change detection method is to find time instants at which the majority of the edge { attributes} in the graph show significant deviation from the recent past. As it can be computationally inefficient to track each edge over time, the authors employ a spectral decomposition approach for this purpose. A positive semi-definite symmetric matrix, $W$, can be decomposed as $W=U\Sigma U^T$, where $U$ is an orthonormal matrix with columns containing the unit eigenvectors of $W$, and $\Sigma$ is a diagonal matrix containing the corresponding eigenvalues \citep{horn2012matrix}. \cite{ide2004eigenspace} extract the principal eigenvector corresponding to the maximum eigenvalue and use it as the representative summary of the graph. The principal eigenvector is a non-negative vector, where larger elements correspond to vertices that have high connectivity or have neighbours with high connectivity in the graph. Hence, \citeauthor{ide2004eigenspace} call this vector the \textit{activity vector}, $\mathbf{u}^t$. This way, the time sequence of graphs can now be represented as a time sequence of activity vectors. At each time instant, $t$, singular value decomposition (SVD) is performed on the matrix whose columns are the activity vectors from the recent past time instants, and the profile behaviour is represented by the principal left singular vector, $\mathbf{r}^{t-1}$, which \cite{ide2004eigenspace} call the \textit{typical activity vector}. Finally a change score, $z^t$, is obtained for each time instant by calculating the dot product between $\mathbf{u}^t$ and $\mathbf{r}^{t-1}$ given by
\begin{equation}\label{eq:ide}
z^t = 1-<\mathbf{u}^t,\mathbf{r}^{t-1}>.
\end{equation}

As $\mathbf{u}^t$ and $\mathbf{r}^{t-1}$ are unit vectors, $z^t$ corresponds to the cosine of the angle between $\mathbf{u}^t$ and  $\mathbf{r}^{t-1}$. Greater deviations from the recent past behaviour correspond to larger angles which give higher change scores. The activity vector at the detected change point is further investigated to detect the vertices that are mostly responsible for the change. 

\begin{figure}
	\centering
		\includegraphics[trim = 0mm 0mm 0mm 8mm, scale=0.5]{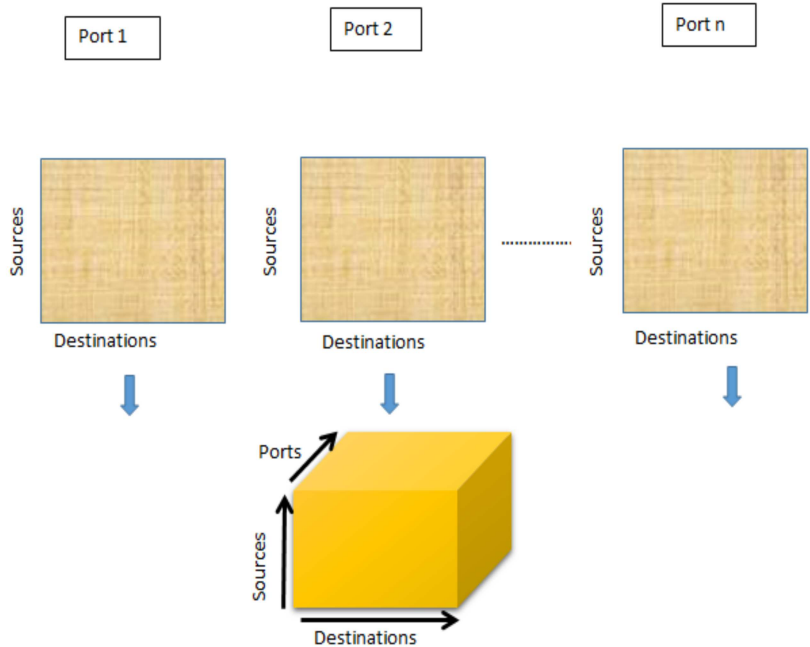}
		\caption{{Representing multiple edge attributes as a 3-D tensor}}
	\label{illustrateTensor}
\end{figure}

In most real-world phenomena, we have multiple sources of information describing different types of relationships associated with entities, which can be captured by a \textit{multi-view network}. For example, we can build a multi-view network over a given set of people based on their relationships over multiple social networking sites such as Facebook, Twitter and Youtube \citep{tang2012community}. A multi-view network is conceptualized as a multi-graph, which is a graph with multiple edge attributes. When a graph has multiple edge attributes, a tensor is preferable to a matrix in order to capture the complex connectivity structure. A tensor is a multidimensional array \footnote{Note that a matrix is a special type of tensor, that is of dimension two.}. There are two ways to represent a dynamic multi-view network as a tensor. First, a dynamic multi-view network can be represented as a time sequence of three-dimensional tensors. Here, the first two dimensions of each tensor denote the vertices, and the third dimension denotes the { categorical edge attributes}. For example, consider a dataset containing port usage in a computer network observed at specific time instants. This dataset may contain the source IP addresses, destination IP addresses, and the port numbers used. Each time instant can then be represented as a tensor, where the first two dimensions represent the sources and destinations, and the third dimension represents the port numbers. { The tensor elements may be weighted or binary. In a weighted tensor, each element quantifies a} communication between the corresponding source and destination via the corresponding port. Figure \ref{illustrateTensor} shows how the sources, destinations, and ports are represented as a three dimensional tensor. The second way is to represent the whole multi-view dynamic network as a four dimensional tensor. The first three dimensions constitute the tensor in the first method, while time is added as the fourth dimension. 

{ As discussed above, a tensor can be used to represent both quantitative and categorical edge attributes pertaining to a dynamic network. Thus, we review some previous tensor-based change detection methods.} The majority of tensor-based change detection methods use tensor decomposition to obtain a low-rank approximation,~$\mathbf{\tilde{A}}$, of the original tensor, $\mathbf{A}$, at each time instant (a detailed review on tensor decomposition methods is provided in \cite{kolda2009tensor}). The reconstruction error,~$e = \parallel \mathbf{A} - \mathbf{\tilde{A}} \parallel_F^2$, measures how well the tensor is approximated at that time instant. \cite{sun2006beyond} decompose each tensor, $\mathbf{A}$, and obtain the reconstruction error,~$e$. If~$e$ is~$2$ standard deviations away from the mean of error values so far, that time instant is characterized as a change point. The tensor-based change detection methods that we find in the literature mainly differ by the technique employed to decompose the tensor. For example, \cite{kolda2008scalable} detect change points by doing Memory Efficient Tucker (MET) decomposition. MET addresses the \textit{immediate blow-up problem} which occurs when the input and output tensors exceed the available memory space during computations for large and sparse graphs.  \cite{papalexakis2012parcube}, on the other hand, employ the PARAFAC tensor decomposition technique \citep{harshman1972determination} in their change detection method. For a comprehensive discussion on tensors and tensor decomposition methods, we direct the reader to \cite{kolda2009tensor}. Furthermore, a collection of tensor-based anomaly detection methods is extensively surveyed in \cite{fanaee2016tensor}.

Calculating similarity between graphs is a well studied problem \citep{faloutsos2009anomaly}, but there are a very few papers that make use of the attributes of vertices { or} edges.  \cite{papadimitriou2010web} introduce several similarity metrics to detect change points in an evolving web network. This network is represented as a time sequence of directed, { attributed} graphs where vertices correspond to active web pages, and { attributed} edges represent the number of hyperlinks between these web pages. \citeauthor{papadimitriou2010web} calculate similarity between consecutive graphs based on the ranks of the vertices and edges. The  rank of a vertex is calculated using its \textit{quality score} obtained from the PageRank method in \cite{eiron2004ranking}. 
%
%
The rank of an edge is calculated by multiplying the weight of a given edge by the quality of the vertices adjacent to it and then normalizing by a constant factor. 
By comparing similarity between all consecutive graphs, a time sequence of similarity measures is obtained. The measure is $0$ for perfect dissimilarity and $1$ for perfect similarity. \citeauthor{papadimitriou2010web} use a simple thresholding algorithm to detect time instants with high dissimilarity. However, time series analysis techniques can be used after obtaining a time sequence of similarity scores as a more sophisticated method to detect change points. For example, autoregressive integrated moving average (ARIMA) method was employed in \cite{pincombe2005anomaly}'s change detection algorithm.

{ \cite{gahrooei2017change} use a probabilistic model based approach to detect global structural changes occurring in a dynamic attributed network. They model each static attributed graph using a generalized linear model (GLM), where the probability of an edge between two vertices is defined as a function of the categorical attributes attached to them. For example, in a social network, these attributes can be the age, location and occupation of the corresponding entities. The GLM assumes each edge to be generated from a probability distribution in the exponential family, such as Bernoulli, Poisson, normal, gamma and so on. Thus, the GLM framework facilitates to model a range of edge attributes such as edge existence, rate of connectivity and volume of connections. The dependency between the time sequence of graphs is captured by constructing a state-space model on the parameters of the GLM, and by using an extended Kalman filter to estimate and update parameters over time. The parameters estimated at $t-1$ are used to predict the graph at $t$ and calculate Pearson residuals. These residuals are monitored over time using an exponentially weighted moving average (EWMA) control chart to detect global structural changes in the stream of graphs.} 

Non parametric probabilistic methods have the flexibility of defining the structure of the model from the data itself without adhering to prior assumptions. One of the most basic non-parametric methods is the histogram. \cite{borges2011anomaly} employ a histogram-based approach to detect change points by studying the distribution of vertex attributes in a dynamic network. At each time instant, eight vertex attributes such as degree, betweenness, closeness, eigenvector centrality and so on are collected for every vertex in the graph. Given a time sequence of graphs, the joint probability distribution of the eight vertex attributes is defined as a histogram, $h_G$. The histogram for the joint vertex attributes up to time instant $t$, $h_{G^{t}}$ is then converted to a density estimate, $\hat{\rho_{G^{t}}}$ using add-one smoothing method. The divergence between the two density estimates,  $\hat{\rho_{G^t}}$ and $\hat{\rho_{G^{t-1}}}$, is then measured using the cross entropy between them. This is defined as the change score for $G^t$. A high divergence score implies significant change in the behaviour of the global structure compared to the past.

{ \cite{mcculloh2011detecting} describe a method based on cumulative sum (CUSUM) control chart to detect changes in a dynamic social network. Changes are detected for graph level measures  (such density, eigenvector centrality) as well as vertex level measures (closeness, betweenness) averaged over the graph. At each time instant, the mean and variance of the measure during recent past is compared with the current measure to calculate the CUSUM’s $C^+$ and $C^-$ statistics. These measures are then compared to a control limit to determine if a change is indicated. Once a change point is detected, the corresponding network is further investigated to identify the responsible entities or connections. The CUSUM chart is well suited for detecting small or gradual changes in an unsupervised environment. However, this approach assumes the network measures to be normally distributed. Thus, the application of this method is limited.}

Change detection methods discussed in Section \ref{Changes in Global Structure} focus on deviations in the behaviour of the overall network. Some changes occurring in the network might not affect the whole network. These changes would go unnoticed, unless the investigation is narrowed down to local structures.

\subsection{\sffamily \large CHANGES IN VERTICES}
\label{Entity Based Methods}

Vertex-based methods are very useful when we want to track the behaviour of important entities in the network. \cite{akoglu2010event} optimize the decomposition-based strategies discussed in \cite{ide2004eigenspace} (Section \ref{Changes in Global Structure}) to detect changes in vertex behaviour. The inputs to their algorithm are twelve vertex attributes observed at each vertex over time.  The whole dataset is first represented as a three-dimensional array, where the three dimensions are the N vertices, F attributes, and $\mathcal{T}$ time instants. To detect changes of vertex $i$, the $\mathcal{T} \times F$ slice of the array corresponding to vertex $i$ is first selected. A window of size $w$ is then defined over F time series. The correlation between each pair of attributes within the window is calculated using Pearson's correlation coefficient, resulting in a correlation matrix for all F attributes. By moving the window to cover all $\mathcal{T}$ time instants, a time sequence of correlation matrices is obtained. Let $u_{i,f}^t$ be the value of attribute $f$ for vertex $i$ at time instant $t$. The $\mathcal{T}\times F$ slice corresponding to vertex $i$ includes the set of time sequences, $\{u_{i,f}^t\}$ for $f\in \{1,\ldots,F\}$ and $t \in \{1,\ldots,\mathcal{T}\}$. Let $C_i^t$ be defined as the matrix calculated at time instant $t$ for vertex $i$, where each element, $[C_i^t]_{f',f''}$, gives the absolute value of the correlation between $u_{i,f'}^t$ and $u_{i,f''}^t$ by
\begin{eqnarray}
[B_i^t]_{f',f''} = \frac{1}{w}\sum_{s=t-w+1}^{t}(u_{i,f'}^s-\bar{u}_{i,f'}^t)(u_{i,f''}^s-\bar{u}_{i,f''}^t),
\end{eqnarray}
where
\begin{equation}
\bar{u}_{i,f'}^t = \frac{1}{w}\sum_{s=t-w+1}^{t}u_{i,f'}^s,
\end{equation}
and
\begin{equation}
[C_i^t]_{f',f''} = \left| \frac{[B_i^t]_{f',f''}}{\sqrt{[B_i^t]_{f',f'}[B_i^t]_{f'',f''}}} \right|. 
\end{equation}

Each element in $C_i$ denotes the strength of the relationship between each pair of attributes. Thus, matrices $C_i^1, \ldots C_i^\mathcal{T}$ are analogous to the weighted adjacency matrices analysed in \cite{ide2004eigenspace} (Section \ref{Changes in Global Structure}). \cite{akoglu2010event} apply \cite{ide2004eigenspace}'s activity vector-based change detection procedure over $C_i^1, \ldots C_i^\mathcal{T}$ to detect changes in vertex $i$'s behaviour. However, for a large dynamic network consisting of a large number of entities, it would not be computationally efficient to detect changes in every vertex's behaviour using this method. Hence, this method is more suitable to track a small number of selected vertices that are identified as interesting in advance.

{ In most scientific studies, data are generated as multivariate time series. Some examples include, climate studies where time varying data are collected on multiple variables such as temperature, precipitation and water discharges \citep{jaruvskova1997some} and studies on financial markets where data on asset returns are observed over time \citep{lavielle2006detection}. Given multivariate time series data, a change in the correlation structure may indicate a change-point in the overall system. For example, neurons in parts of the brain show extreme synchronisation at the time of occurrence of a seizure \citep{jeremy2009sources}. Thus, many researches have been conducted in detecting correlation changes in multivariate time series data \citep{aue2009break,muller2005detection,bulteel2014decon,cabrieto2017detecting}. As discussed previously in Section \ref{Entity Based Methods}, \cite{akoglu2010event} obtain a  time sequence of graphs from multivariate time series data and detect changes in their correlation structure. We believe that it would be worthwhile to investigate the suitability of such a graph-based approach to tackle the change-point detection problem in multivariate time series data. 
	
In \cite{jiang2015link}, the similarity between the categorical vertex attributes in the graph are used to calculate a weighted adjacency matrix, where the weights can be the correlation coefficient or the cosine similarity between the vector of attributes attached to the two vertices forming an edge. However, \cite{jiang2015link}'s goal is edge prediction. Nevertheless, this idea can also be utilized for the purpose of change detection. For example, it is possible to combine the weighted adjacency matrix obtained using categorical vertex attributes with the weighted adjacency matrix obtained using quantitative vertex attributes \citep{akoglu2010event} or edge attributes \citep{ide2004eigenspace} using a tensor (see Section \ref{Changes in Global Structure} for a detailed discussion on tensors). It would be interesting to investigate whether the addition of categorical vertex attributes can improve change detection performance.}

\cite{rossi2013modeling} use multiple vertex attributes to detect vertices with unusual transitions in their behaviour. Each time instant is first represented as a vertex-by-attribute matrix, $W^t \in \mathbb{R}^{n \times f}$. These $f$ attributes for the $n$ vertices are calculated from the graph by using Rolx method discussed in \cite{henderson2012rolx}. Using Non-negative Matrix Factorization (NMF) \citep{henderson2012rolx}, each $W^t$ is then decomposed as $W^t=Z^t \times F$, where $Z^t \in \mathbb{R}^{n \times r}$ and $F \in \mathbb{R}^{r \times f} $ . The $r$ columns of matrix $Z^t$ contain the $r$ latent factors for the $f$ attributes in $W^t$, which are described as \textit{roles} in the paper. Thus, the $n$ rows of $Z^t$ describe the membership of each vertex corresponding to the $r$ roles at time instant $t$. For each consecutive pair of time instants, the role transition model for a particular vertex, $i$, is estimated using NMF as $\hat{Z^t_i}=T_i^t Z^{t-1}_i$. Based on the estimated role membership matrix, $\hat{Z^t_i}$, the change score for vertex $i$ is calculated as $\parallel \hat{Z^t_i}-Z^t_i \parallel$. Using this method, significant changes in a vertex's role transition behaviour can be detected. The drawback of being restricted to compare only consecutive time instants is removed by employing a stacked transition model. This model generalizes the former concept to compare $k$ previous time instants from recent past with the current time instant $t$.  

\subsection{\sffamily \large CHANGES IN EDGES}
\label{Edge Based Methods}

These methods aim to detect individual edges or groups of edges that show changes in their behaviour. Some examples of edge-based changes include the appearance of an edge between two vertices in the graph which represents entities that are very unlikely to interact with each other, or a large increase or decrease in the weight of an edge compared to the recent past. For example, in an author conference network, we may observe an author who usually publishes papers only in conference proceedings related to a particular research area. However, at a certain point in time, he switches to a new research area and publishes a paper in a related conference proceedings. This forms an irregular connection with respect to his past connections in the network \citep{koutra2012tensorsplat}.

\cite{sricharan2014localizing} use a method based on \textit{commute time distances} to detect changes involving edges. Given a graph, the commute time distance between vertex $i$ and vertex $j$ is the expected return path length between vertex $i$ and vertex $j$ \citep{asz80random}. A fast and reliable method was introduced by \cite{khoa2012large} to calculate commute time distances between vertices in large graphs. \cite{sricharan2014localizing} adopt this method to detect edges that undergo changes in a time sequence of { attributed} graphs with weighted edges. Their change detection method, CAD (commute time-based anomaly detection) measures change in the commute time distance between two vertices in a graph inorder to detect changes in the associated edges. First, the commute time distances between every pair of vertices in a graph are calculated. The change-score, $Z^{t}_{i,j}$, for an edge between vertex $i$ and vertex $j$ is then given by
\begin{equation}\label{eq:sricharan}
Z^{t}_{i,j}=|W^{t}_{i,j}-W^{t-1}_{i,j}|\times |C^{t}_{i,j}-C^{t-1}_{i,j}|, 
\end{equation}
where $W^t$ is the weighted adjacency matrix and $C^t$ is the commute time distance matrix. A change in the edge weight between vertices, $i$ and $j$, in turn affects the commute time distances between the neighbours of vertex $i$ and vertex $j$. Measuring the change using $C^{t}$ and $C^{t-1}$ alone has been shown to result in many high change scores even for unchanged edges. Thus, \cite{sricharan2014localizing} include the term, $|W^{t}_{i,j}-W^{t-1}_{i,j}|$, in Equation \ref{eq:sricharan} to address this problem of false detections. According to the experimental results in \cite{sricharan2014localizing}, CAD shows good performance  in detecting changing relationships in many real-world networks. However, as CAD can only perform pairwise comparisons between consecutive graphs, it cannot be applied when several graphs from the recent past are required to characterize usual behaviour.

\cite{heard2010bayesian} detect suspicious connections in a communication network with attributed edges using a two stage Bayesian approach. The number of communications between two vertices in a given time instant is taken as the edge attribute of the corresponding network. The edges in each time instant are modelled as a discrete time counting process. Based on the distribution derived from previous time instants, edges in the current time instant are assigned a \textit{p}-value to quantify change. The vertices connected to the edges going through change are also detected. Finally, substructures formed by the detected vertices are further examined to identify suspicious regions in the network. \cite{turcotte2014intended} in their work extended \citeauthor{heard2010bayesian}'s work by adjusting the model to incorporate seasonality, which is a common characteristic in time varying communication networks.

In Section \ref{Changes in Global Structure}, we discuss how the reconstruction error of a tensor can be employed to detect changes in a dynamic multi-view network. Let $\mathbf{A}^t$ be the tensor at time instant $t$ at which a global structural change has occurred. The search can be further narrowed down to detect edges that have changed their behaviour the most. This can be achieved by examining the elements of the residual tensor, $\mathbf{R}^t = |\mathbf{A}^t - \mathbf{\tilde{A}}^t|$ \citep{ranshous2015anomaly}. The elements in $\mathbf{R}^t$ that show the highest scores correspond to edges that have changed most. 

\subsection{\sffamily \large CHANGES IN SUBGRAPHS}
\label{Sub-graph Based Methods}

In some scenarios, unusual activity spreads throughout a certain neighbourhood in a network and prevails for some time period. For example, in a road network that consists of street intersections connected by road segments, an accident may suddenly increase the traffic density in road segments near to the accident. Such a change is reflected as a change in the behaviour of some vertices and edges in the corresponding graph, where road segments are denoted as edges and street intersections are denoted as vertices. In this case, rather than observing the behaviour of individual vertices or edges, it is more useful to keep track of a \textit{local region} in the graph. A local region is a subgraph formed by a set of vertices that lie in close proximity to each other \citep{ide2009proximity}, such as, a set of vertices in a $k$-\textit{path} (discussed later in the section) \citep{neil2013scan}, or a \textit{clique}\footnote{A \textit{clique} is a set of vertices, that are pairwise connected to each other.} \citep{chen2012community}. 

\begin{figure}[h]
	\centering
	\includegraphics[trim = 0mm 0mm 0mm 0mm, scale=0.5]{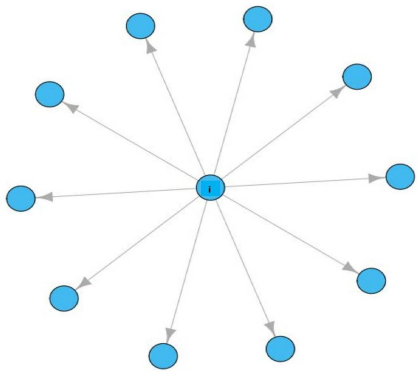}
	\caption{{The star shaped subgraph in \cite{priebe2005scan} for a first order neighbourhood (taken from \cite{neil2013scan} with permission)}. \label{IllustratestarScan}}
\end{figure}

The concept of a \textit{scan statistic} originated in spatial statistics and image analysis to detect regions of unusual activity \citep{marchette2012scan}. Scan statistics were then adopted in the area of network analysis to detect anomalous regions in a graph. Usually, a subgraph is defined centred at each vertex in the graph and a statistic is calculated. The subgraphs that show unusual values for this statistic are highlighted as anomalous. \cite{priebe2005scan} use a scan statistics approach for change detection by detecting a subgraph that shows an unusual increase in interactions with respect to the recent past. The authors apply this method to a time sequence of graphs extracted from the Enron e-mail corpus\footnote{\url{https://www.cs.cmu.edu/~./enron/}}. Each graph represents the email communications between a group of Enron executives during a week, and contains directed, binary edges. The $k$\textit{th order neighbourhood} of vertex $i$ is the set of all vertices that lie within a shortest path length of at most $k$ in the graph. The \textit{locality statistic} for vertex $i$, $\Psi_{k,i}^t$, at time instant $t$ is defined as the size of its $k^{th}$-order neighbourhood. { The locality statistic, $\Psi_{k,i}^t$, can also be considered as an attribute attached to a vertex at a given time instant}. 
The maximum of the locality statistics, $\max_i \Psi_{k,i}^t$, is identified as the \textit{scan statistic} at that time instant. By moving a window of length $w$ over the time sequence of scan statistics, a change score is obtained at each time instant. Time instants that give high change scores are detected as change points. For a given change point, $t$, the vertices corresponding to $\max_i \Psi_{k,i}^t$ are regarded as the vertices responsible for the detected change. When $k=1$, the subgraph defining $\Psi_{k,i}^t$ in \cite{priebe2005scan} resembles a star that is centred at vertex $i$ (Figure \ref{IllustratestarScan}). 

\begin{figure}[h]
	\centering
	\includegraphics[trim = 0mm 0mm 0mm 0mm, scale=0.7]{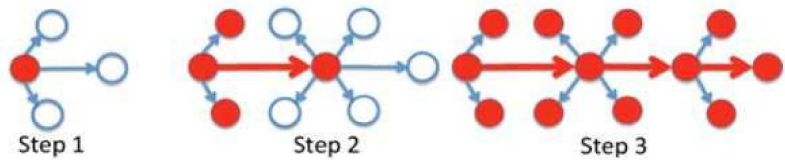}
	\caption{{The traversal behaviour of an intruder in a computer network (taken from \cite{neil2013scan} with permission)}. \label{IllustrateTraverse}}
\end{figure}

Computer networks are vulnerable to attacks by intruders. One common type of intruder behaviour is \textit{intruder traversal} \citep{neil2013scan}. For example, an intruder sends an email  containing a link to a malicious web site to a set of users. When a user clicks on the link, the user's computer gets infected and the intruder gains access to that computer. From this computer the intruder then compromises to another computer in a similar manner. In this way, the intruder traverses the whole computer network, accessing valuable data at each infected computer. Figure \ref{IllustrateTraverse} illustrates the traversal behaviour of an intruder in a network. 
\cite{neil2013scan} introduce a $k$-\textit{path} shaped subgraph to detect intruder traversal in a computer network. A $k$-path is a directed subgraph, where both its size and diameter are equal to $k$ \citep{kolaczyk2009statistical}. \cite{neil2013scan} obtain scan statistics by fitting a Hidden Markov Model (HMM) on the edges of each $k$-path in the graph. Similar to the procedure followed in \cite{priebe2005scan}, $k$-paths that show different behaviour compared to their recent past are detected using a moving window approach.

A chatter anomaly is a group of vertices that show increased number of connections compared to their past behaviour. \cite{pao2011statistical} compare the power of several statistical tests in detecting a chatter anomaly appearing in an Erd\H{o}s R\`{e}nyi random graph \citep{erdHos1959random}.  They showed that none of the considered global (size,order) as well as local (degree, size of $k$th order neighbourhood) graph attributes produced a uniformly powerful test for the purpose of the detection of the chatter anomaly. A recent study by \cite{park2013anomaly} utilizes these graph attributes simultaneously to detect change points caused by a chatter anomaly in a dynamic network. The graphs are assumed to be generated from a latent process model \citep{lee2011latent} (detailed discussed in Section \ref{Discussion}). The new statistic the authors derive, is a linear combination of the nine graph attributes discussed in \cite{pao2011statistical}'s work. \cite{tang2013attribute} investigate the same test statistic for change detection on an attributed random graph, where the edges are annotated with a single categorical attribute.

Although not directly targeted at change detection, the { `Timecrunch'} algorithm introduced in \cite{shah2015timecrunch} employ an information theoretic approach to understand the behaviour of a particular subgraph observed at each time instant. They assume that the most common types of subgraphs that can be found in real world networks are stars, full and near cliques, full and near bipartite cores and chains \citep{kleinberg1999web}. Each time instant is analysed to find the above mentioned subgraph types. Using a Minimum Description Length (MDL) principle, the temporal behaviour of these subgraphs is inferred. Two subgraphs in different time instants are identified as the same type, only if there is a substantial overlap in the set of vertices comprising them. Subgraphs that appear only at one time instant are classified as `Oneshot'. A subgraph that appears over a period of consecutive time instants is called as `Ranged'. Some subgraphs appear and disappear in random time instants, these are characterized as `Flickering'. Subgraphs that appear at fixed time intervals showing a seasonal behaviour are characterized as `Periodic'. `Constant' subgraphs are the ones that are found to be present across all time instants.

\subsubsection*{\sffamily \normalsize Communities}
\label{Community based methods}

A \textit{community} is a special type of subgraph. There exist several definitions of a community in the network analysis literature. In some publications that use probabilistic model-based approaches to detect community structure in graphs \citep{holland1983stochastic,aicher2014learning}, a community is defined as a set of vertices that are stochastically equivalent. In other publications, a community is defined as a dense subgraph. Some examples along this line of research include, community detection through spectral clustering \citep{ng2001spectral}, hierarchical clustering \citep{newman2004detecting} and maximization of the so called ``Newman–Girvan'' modularity function \citep{newman2004detecting}. In all the latter mentioned publications, a community consists of a group of vertices that have higher number of edges within the group than between the group and the rest of the graph. For example, in a social network, a community may correspond to a board of directors in a company who have stronger connections among themselves than with others. 

\begin{figure}[h]
	\centering
	\includegraphics[trim = 0mm 0mm 0mm 0mm, scale=0.8]{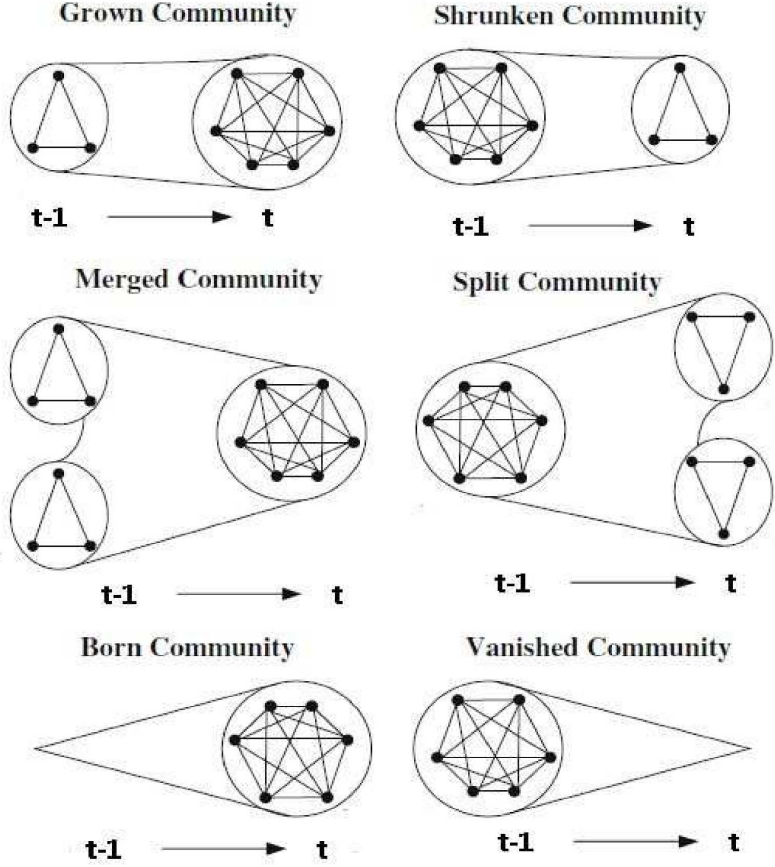}
	\caption{{Possible changes occurring in community structure (taken from \cite{chen2012community} with permission)}. \label{Illustratemerge}}
\end{figure}

Communities in real-world networks are usually stable over time \citep{koujaku2013structrual}. Thus, deformation in community structure is a good indication of a change occurring in the network.  \cite{chen2012community} { define a community as a dense subgraph and discuss} six basic changes that can occur in community structure of a graph (Figure \ref{Illustratemerge}):
\begin{enumerate}
	\item \textit{Grown community}\\
	Vertices get added into a smaller community, causing the community to grow into a larger community. The size of the community at time instant $t$ is greater than its size at $t-1$.
	\item \textit{Shrunken community}\\
	Community members leave a community, causing the community to become smaller. The size of the community at time instant $t$ is less than its size at $t-1$.
	\item \textit{Merged community}\\
	At time instant $t-1$, there were two communities. These two communities join together and form a single larger community at time instant $t$. 
	\item \textit{Split community}\\
	One community at time instant $t-1$ divides into two communities at time instant $t$.
	\item \textit{Born community}\\
	The edges between a set of vertices increase and form a community at time instant $t$. This scenario uses the same notion as the \textit{form} scenario discussed in \cite{peel2015detecting}.
	\item \textit{Vanished community}\\
	A community that prevailed at time instant $t-1$ disappears at time instant $t$ because relationships between members cease to exist. The same change is also referred to as \textit{fragment\textit{}} in \cite{peel2015detecting}.
	
\end{enumerate}

\begin{figure}[h]
	\centering
	\includegraphics[trim = 0mm 0mm 0mm 0mm, scale=1]{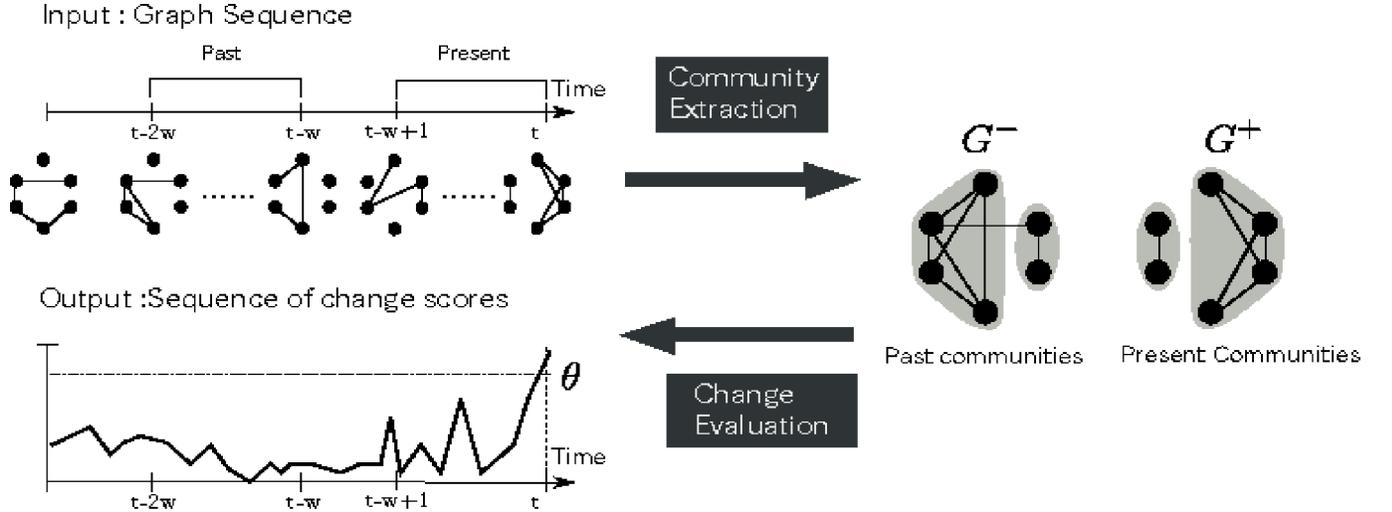}
	\caption{{Illustration of \cite{koujaku2015community}'s change detection framework with double sliding windows (taken from \cite{koujaku2015community} with permission)}. \label{Illustrate_loujaku}}
\end{figure}

\cite{koujaku2015community} perform experiments to detect the above-mentioned six types of changes in community structure. Their method employs two moving windows for change score calculation. At each time instant, the first window contains time instants from $t-w+1$ to $t$ and the second window contains time instants from $t+1$ to $t+w$. Let $G^-$ be the graph constructed from the time instants in the first window and let $G^+$ be the graph constructed from the time instants in the second window. A set of communities, $\mathcal{U}^-$, and another set,   $\mathcal{U}^+$, are extracted\footnote{\cite{koujaku2015community} explore three algorithms, (i) graph scan \citep{wang2008spatial}, (ii) extraction method \citep{zhao2011community}, and (iii) $\rho-$dense core \citep{koujaku2013structrual}, for community extraction.} from $G^-$ and $G^+$, respectively. A change score, $z^t$, is then calculated by using the \textit{Variational Information} (VI) \citep{meilua2003comparing} between $\mathcal{U}^+$ and $\mathcal{U}^-$. This is given by
\begin{eqnarray}
z^t & = & \text{VI}(\mathcal{U}^+,\mathcal{U}^-) \nonumber \\
& = & H(\mathcal{U}^+) + H(\mathcal{U}^-) - 2I(\mathcal{U}^+,\mathcal{U}^-),
\end{eqnarray}
where $H$ and $I$ are the \textit{entropy} and \textit{mutual entropy} defined as
{

\begin{equation}
H(\mathcal{U}) = \sum_{V_1 \in \mathcal{U}} \frac{|V_1|}{|V|} \log \left( \frac{|V_1|}{|V|}\right),
\end{equation}
\begin{equation}
I(\mathcal{U^+},\mathcal{U^-})= \sum_{V_1\in \mathcal{U^+}}\sum_{V_2\in \mathcal{U^-}} \frac{|V_1 \cap V_2 |}{|V|}\log \left( \frac{|V||V_1 \cap V_2 |}{|V_1||V_2|}   \right).
\end{equation}

}
Finally, once a time sequence of change scores is obtained, a threshold, $\theta$, is defined and time instants with change scores exceeding this threshold are detected as change points. Figure \ref{Illustrate_loujaku} is an illustration of \cite{koujaku2015community}'s change detection procedure. 

\begin{figure}[h]
	\centering
	\includegraphics[trim = 0mm 0mm 0mm 0mm, scale=0.7]{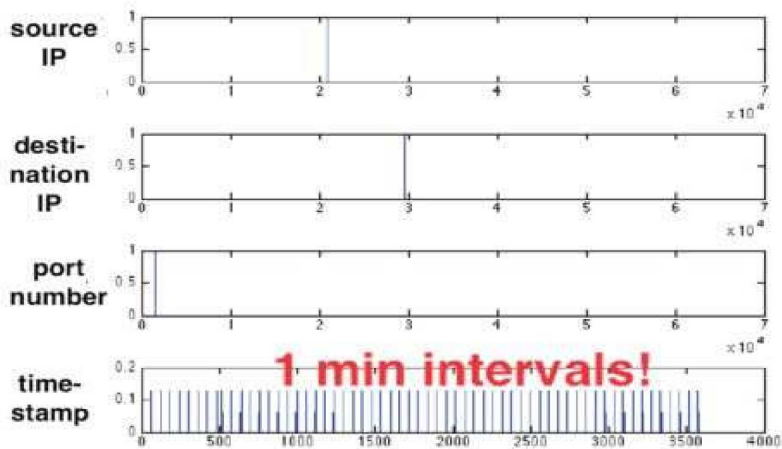}
	\caption{{Bot-attack like behaviour (taken from \cite{koutra2012tensorsplat} with permission)}.  \label{Illustrate_botAttack}}
\end{figure}

\begin{figure}[htbp]
	\centering
	\includegraphics[trim = 0mm 0mm 0mm 0mm, scale=0.7]{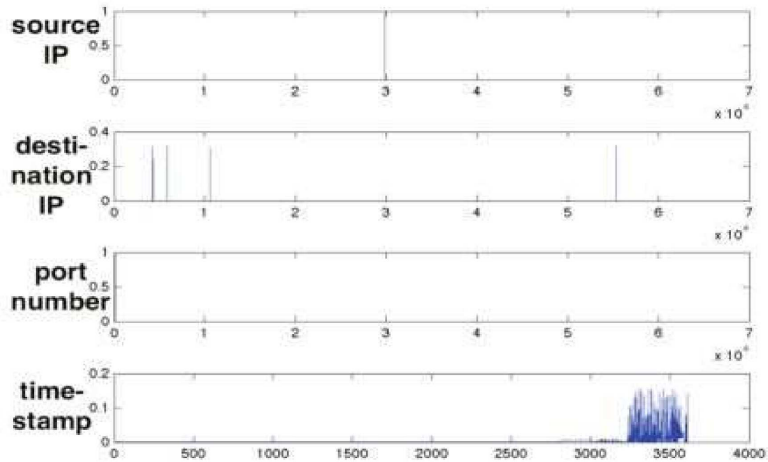}
	\caption{{Traffic caused by human activity (taken from \cite{koutra2012tensorsplat} with permission)}. \label{Illustrate_human}}
\end{figure}

In Section \ref{Changes in Global Structure}, we discuss how we can represent a dynamic network with multiple edge attributes as a four-dimensional tensor. \cite{koutra2012tensorsplat} represent the traffic measurements of a computer network as an ${n \times m \times f \times t}$ tensor, $\mathbf{A}$, where the four dimensions represent source IP addresses, destination IP addresses, port numbers of the connections and time instants, respectively. They then do PARAFAC decomposition \citep{harshman1970foundations} and obtain  factor matrices,
$U\in \mathbb{R}^{n \times f'}, V\in \mathbb{R}^{m \times f'}, W\in \mathbb{R}^{f \times f'},$ and $X\in \mathbb{R}^{t \times f'}$, where $f' < \min \{n,m,f,t\}$. Each column of $U$ represents a community of sources (source IP addresses) { that have higher number of connections within themselves}. Each column of $V$ represents a community of destinations (destination IP addresses). The columns of $W$ correspond to the port numbers of the connections established between communities represented by the respective columns in $U$ and $V$, while the columns of $X$ give temporal profiles of these communities. \citeauthor{koutra2012tensorsplat} select the $f'$ most important columns of these factor matrices, track them inorder to investigate changes in user behaviour and to detect intruder attacks in the computer network. For example, Figure \ref{Illustrate_botAttack} shows a plot of the same columns taken from each of the factor matrices, $U$,$V$, $W$ and $X$. We observe suspicious behaviour on port $1544$. The connection on port $1544$ by a particular source IP address to a particular destination IP address is established periodically with perfectly evenly spaced spikes of activity. This type of behaviour is called a \textit{bot attack}. Intruders in computer networks often exhibit this type of behaviour \citep{koutra2012tensorsplat}. In Figure \ref{Illustrate_human}, from another common column taken from each of the four factor matrices, we observe an overwhelming  burst of traffic concentrated in a certain time interval on port $80$ by a particular IP address connected to several destination IP addresses. This type of behaviour resembles a human browsing several websites during a given time interval. 
\cite{koutra2012tensorsplat}'s tensor-based change detection method is called \textit{tensor-splat}. Tensor-splat is used for detecting changes in many real-world networks. For example, the DBLP dataset\footnote{\url{http://dblp.uni-trier.de/xml/}} consists of a large collection of papers that appeared at different conferences written by a number of authors over several years. \cite{koutra2012tensorsplat} build a tensor with dimensions, ${418000 \times 3.5000 \times 49}$, with the three dimensions representing authors, conferences, and years, respectively. When the tensor is decomposed using PARAFAC as described earlier, the communities indicated by the factor matrix representing authors, signify groups of authors who come from similar research areas. Using tensor-splat, \citeauthor{koutra2012tensorsplat} detect \textit{bridge authors} who gradually changed their research area over time.

\cite{peel2015detecting} discuss a parametric probabilistic approach to change detection. They assume each graph in the time sequence to be generated from a generalized hierarchical random graph model (GHRG), which is an improved version of the hierarchical random graph model (HRG) discussed in \cite{clauset2007structural}. { Considering a community as a group of vertices that are stochastically equivalent}, this model has the ability to capture assortative and disassortative community structure, while at the same time providing interpretable results for different types of networks. \citeauthor{peel2015detecting} detect change points as the time instants where the model parameters differ significantly from the expected model. By defining a window, a posterior Bayes factor is calculated between every pair of consecutive graphs to decide whether there is a parameter change or not. The maximum Bayes factor obtained within a window is the test statistic.  This way, a time sequence of test statistics is obtained by moving the window across all graphs in the sequence. This sequence of test statistics is then input into a threshold calculation algorithm to detect the change points. The probabilistic model in \cite{peel2015detecting} can easily be replaced with any random graph model. Thus, any model that can handle the vertex or edge attribute information can replace the GHRG model. For example we can employ the Weighted Stochastic Block Model (WSBM) \citep{aicher2014learning}, that models both edge presence and edge weights (categorical or quantitative) of the graph. 

\section{\sffamily \Large DISCUSSION}
\label{Discussion}

{ As there are various types of change detection methods available, it is difficult to decide which ones would be useful for a given scenario. Selecting a method can be done by evaluating it's performance. One may apply the method to a real world scenario and check whether the results agree with their ground truth information.} Several dynamic datasets are available online { for this purpose}. The Enron { email} dataset (\url{http://www.cs.cmu.edu/~enron/}) { provides information regarding emails exchanged in the Enron company. Enron was declared bankrupt in 2002 due to accountancy
fraud activities by the top level executives. Since the Enron dataset is rich with ground truth information, it is used to evaluate the performance of many change detection methods \citep{azarnoush2016monitoring,li2016social,peel2015detecting,rossi2013modeling,tang2008community}.} Another popular dynamic network dataset is the Dblp dataset (\url{http://dblp.uni-trier.de/db/}).{  It contains bibliographic information in the area of computer science. \cite{koutra2012tensorsplat}, \cite{yu2014glad}, \cite{sun2006beyond} and \cite{ji2013incremental} are some of the papers that use the Dblp dataset in their experiments. Their discussions are helpful in evaluating the outcomes of a new change detection method. Furthermore,} \cite{ranshous2015anomaly} in their review paper, provide a comprehensive collection of publicly available datasets. In addition to these resources, we find some more interesting dynamic network datasets in \url{<http://projects.csail.mit.edu/dnd/>}. However, majority of these datasets do not contain multiple attribute information on vertices and edges. We direct the interested reader to \url{<http://leitang.net/heterogeneous_network.html>}, which provides access to three datasets, Enron, Dblp and You-tube that are rich with attribute information and also contain references to past publications { that have analysed these datasets}. { Moreover}, the `High School Contact and Friendship' dataset available at \url{<http://www.sociopatterns.org/datasets/>} contains heterogeneous edge information that can be useful in obtaining an attributed dynamic network. 

Another way to evaluate the quality of a method is by applying it to a simulated network. \cite{leskovec2005realistic} proposed a model that can generate a network with several real-world graph properties such as `preferential attachment', `shrinking diameter' and `densification power law characteristic'. We refer the reader to \cite{aggarwal2014evolutionary} and the references therein for a detailed description about these characteristics. The graph at time $t$ is generated by the Kronecker product of the binary adjacency matrix at time $t-1$ by itself. A certain drawback of this method is the result of large multiplicities on the edges due to the Kronecker product between binary matrices. \citeauthor{leskovec2005realistic} call this as the `staircase effect'. This problem has been resolved by the use of Stochastic Kronecker graphs. This produces a time sequence of weighted graphs where the weight of an edge is the probability of its existence. \cite{akoglu2008rtm} extend the above mentioned Kronecker recursion method to a tensor-based method. Initially the dynamic network is represented as a 3-dimensional tensor where the first and second dimensions represent the adjacency matrix of the graph and the third dimension represents time. Then the initial tensor is multiplied $k$ times  recursively by itself to obtain the final tensor that represents the data. Empirical results in \cite{akoglu2008rtm} show how the graphs generated by this proposed network generator successfully mimic real world properties in weighted and unweighted dynamic networks.

Some authors apply the concepts of latent position models  \citep{hoff2002latent}  to generate a time series of graphs. Latent position models are a group of random graph models that characterize the structure of the graph using latent vectors associated with the vertices. \cite{lee2011latent}, propose the \textit{latent process model} to generate a time sequence of graphs with weighted edges. This model is developed based on the concept of the Random Dot Product Graph (RDPG) model \citep{nickel2007random}. The RDPG model is a special type of latent position model. Here, the probability of each edge in the graph is given by the dot product of the latent vectors associated with the two vertices forming the edge. In the latent process model, the weighted edges are a function of a continuous-time finite-state stochastic process emanating from the vertices. Using a time sequence of networks generated from this model, \cite{park2013anomaly}  detect change points caused by chatter anomalies (see section \ref{Entity Based Methods}) appearing in time. \cite{wang2014locality} also detected chatter anomalies by generating a dynamic network according to a time varying Stochastic Block Model (SBM) \citep{holland1983stochastic}. The SBM is a random graph model that groups the vertices in the graph into several blocks, and the probability of an edge between two vertices depend mainly on their block membership. For each vertex in the graph, \citeauthor{wang2014locality} generate a latent vector using a Dirichlet distribution. Similar to an RDPG model, the probability of an edge between two vertices is then defined as the dot product of the corresponding latent vectors. By tuning the parameters of the Dirichlet distribution, the authors generate graphs from an SBM \citep{sussman2012consistent}. To get a general understanding about the underlying mechanisms of each of these random graph models, we direct the reader to \cite{goldenberg2010survey} and \cite{salter2012review}.

\section{\sffamily \Large CONCLUSION}
\label{Conclusion}

In this paper, we review several change detection methods for dynamic attributed networks. We mainly categorize these approaches based on the structure level of the graph investigated to quantify the change. We also provide useful resources for dynamic network datasets. Finally, we review different synthetic network generating mechanisms for the interested reader.

The field of change detection in dynamic networks is a new and growing area. However, there are only a handful of publications that exploit vertex and edge attribute information. Addition of attribute information result in high dimensional data, that complicate the detection procedure in a dynamic scenario. However, due to its huge applicability to many real world situations, the popularity of this research area is rapidly growing. There is a high demand for scalable algorithms that can efficiently analyse this type of data (Big Data). Throughout this review paper, we guide the reader to possible paths of exploration in this direction.

\section*{\sffamily \Large ACKNOWLEDGEMENTS}
The author extends her gratitude to Dr. Dominic Lee, Prof. Elena Moltchanova, Dr. Jeanette McLeod and Mr. Roger Jarquin from the School of Mathematics and Statistics, University of Canterbury, New Zealand, for providing insight and expertise that greatly assisted this work.



\bibliography{Bibliography}

\begin{thebibliography}{}

\bibitem[Aggarwal and Subbian, 2014]{aggarwal2014evolutionary}
Aggarwal, C. and Subbian, K. (2014).
\newblock Evolutionary network analysis: A survey.
\newblock {\em ACM Computing Surveys (CSUR)}, 47(1):10.

\bibitem[Aicher et~al., 2014]{aicher2014learning}
Aicher, C., Jacobs, A.~Z., and Clauset, A. (2014).
\newblock Learning latent block structure in weighted networks.
\newblock {\em arXiv preprint arXiv:1404.0431}.

\bibitem[Akoglu and Faloutsos, 2010]{akoglu2010event}
Akoglu, L. and Faloutsos, C. (2010).
\newblock Event detection in time series of mobile communication graphs.
\newblock In {\em Army Science Conference}, pages 77--79.

\bibitem[Akoglu et~al., 2008]{akoglu2008rtm}
Akoglu, L., McGlohon, M., and Faloutsos, C. (2008).
\newblock Rtm: Laws and a recursive generator for weighted time-evolving
  graphs.
\newblock In {\em Data Mining, 2008. ICDM'08. Eighth IEEE International
  Conference on}, pages 701--706. IEEE.

\bibitem[Akoglu et~al., 2015]{akoglu2015graph}
Akoglu, L., Tong, H., and Koutra, D. (2015).
\newblock Graph based anomaly detection and description: a survey.
\newblock {\em Data Mining and Knowledge Discovery}, 29(3):626--688.

\bibitem[Aue et~al., 2009]{aue2009break}
Aue, A., H{\"o}rmann, S., Horv{\'a}th, L., Reimherr, M., et~al. (2009).
\newblock Break detection in the covariance structure of multivariate time
  series models.
\newblock {\em The Annals of Statistics}, 37(6B):4046--4087.

\bibitem[Azarnoush et~al., 2016]{azarnoush2016monitoring}
Azarnoush, B., Paynabar, K., Bekki, J., and Runger, G. (2016).
\newblock Monitoring temporal homogeneity in attributed network streams.
\newblock {\em Journal of Quality Technology}, 48(1):28--43.

\bibitem[Borges et~al., 2011]{borges2011anomaly}
Borges, N., Coppersmith, G., Meyer, G.~G., Priebe, C.~E., et~al. (2011).
\newblock Anomaly detection for random graphs using distributions of vertex
  invariants.
\newblock In {\em Information Sciences and Systems (CISS), 2011 45th Annual
  Conference on}, pages 1--6. IEEE.

\bibitem[Bulteel et~al., 2014]{bulteel2014decon}
Bulteel, K., Ceulemans, E., Thompson, R.~J., Waugh, C.~E., Gotlib, I.~H.,
  Tuerlinckx, F., and Kuppens, P. (2014).
\newblock Decon: A tool to detect emotional concordance in multivariate time
  series data of emotional responding.
\newblock {\em Biological psychology}, 98:29--42.

\bibitem[Cabrieto et~al., 2017]{cabrieto2017detecting}
Cabrieto, J., Tuerlinckx, F., Kuppens, P., Grassmann, M., and Ceulemans, E.
  (2017).
\newblock Detecting correlation changes in multivariate time series: A
  comparison of four non-parametric change point detection methods.
\newblock {\em Behavior research methods}, 49(3):988--1005.

\bibitem[Chen et~al., 2012]{chen2012community}
Chen, Z., Hendrix, W., and Samatova, N.~F. (2012).
\newblock Community-based anomaly detection in evolutionary networks.
\newblock {\em Journal of Intelligent Information Systems}, 39(1):59--85.

\bibitem[Cheng et~al., 2008]{cheng2008robust}
Cheng, H., Tan, P.-N., Potter, C., and Klooster, S. (2008).
\newblock A robust graph-based algorithm for detection and characterization of
  anomalies in noisy multivariate time series.
\newblock In {\em 2008 IEEE International Conference on Data Mining Workshops},
  pages 349--358. IEEE.

\bibitem[Clauset et~al., 2007]{clauset2007structural}
Clauset, A., Moore, C., and Newman, M.~E. (2007).
\newblock Structural inference of hierarchies in networks.
\newblock In {\em Statistical network analysis: models, issues, and new
  directions}, pages 1--13. Springer.

\bibitem[Durante and Dunson, 2014]{durante2014bayesian}
Durante, D. and Dunson, D.~B. (2014).
\newblock Bayesian dynamic financial networks with time-varying predictors.
\newblock {\em Statistics \& Probability Letters}, 93:19--26.

\bibitem[Durante et~al., 2014]{durante2014nonparametric}
Durante, D., Dunson, D.~B., and Vogelstein, J.~T. (2014).
\newblock Nonparametric bayes modeling of populations of networks.
\newblock {\em arXiv preprint arXiv:1406.7851}.

\bibitem[Eiron et~al., 2004]{eiron2004ranking}
Eiron, N., McCurley, K.~S., and Tomlin, J.~A. (2004).
\newblock Ranking the web frontier.
\newblock In {\em Proceedings of the 13th international conference on World
  Wide Web}, pages 309--318. ACM.

\bibitem[Erd{\H{o}}s and R{\'e}nyi, 1959]{erdHos1959random}
Erd{\H{o}}s, P. and R{\'e}nyi, A. (1959).
\newblock On random graphs.
\newblock {\em Publicationes Mathematicae Debrecen}, 6:290--297.

\bibitem[Faloutsos, 2009]{faloutsos2009anomaly}
Faloutsos, L. A. M. M.~C. (2009).
\newblock Anomaly detection in large graphs.

\bibitem[Fanaee-T and Gama, 2016]{fanaee2016tensor}
Fanaee-T, H. and Gama, J. (2016).
\newblock Tensor-based anomaly detection: An interdisciplinary survey.
\newblock {\em Knowledge-Based Systems}, 98:130--147.

\bibitem[Fayyad et~al., 1996]{fayyad1996data}
Fayyad, U., Piatetsky-Shapiro, G., and Smyth, P. (1996).
\newblock From data mining to knowledge discovery in databases.
\newblock {\em AI magazine}, 17(3):37.

\bibitem[Gahrooei and Paynabar, 2017]{gahrooei2017change}
Gahrooei, M.~R. and Paynabar, K. (2017).
\newblock Change detection in a dynamic stream of attributed networks.
\newblock {\em arXiv preprint arXiv:1711.04441}.

\bibitem[Goldenberg et~al., 2010]{goldenberg2010survey}
Goldenberg, A., Zheng, A.~X., Fienberg, S.~E., and Airoldi, E.~M. (2010).
\newblock A survey of statistical network models.
\newblock {\em Foundations and Trends{\textregistered} in Machine Learning},
  2(2):129--233.

\bibitem[Harshman, 1970]{harshman1970foundations}
Harshman, R.~A. (1970).
\newblock Foundations of the parafac procedure: Models and conditions for an"
  explanatory" multi-modal factor analysis.

\bibitem[Harshman, 1972]{harshman1972determination}
Harshman, R.~A. (1972).
\newblock Determination and proof of minimum uniqueness conditions for
  parafac1.
\newblock {\em UCLA Working Papers in phonetics}, 22(111-117):3.

\bibitem[Heard et~al., 2010]{heard2010bayesian}
Heard, N.~A., Weston, D.~J., Platanioti, K., Hand, D.~J., et~al. (2010).
\newblock Bayesian anomaly detection methods for social networks.
\newblock {\em The Annals of Applied Statistics}, 4(2):645--662.

\bibitem[Henderson et~al., 2012]{henderson2012rolx}
Henderson, K., Gallagher, B., Eliassi-Rad, T., Tong, H., Basu, S., Akoglu, L.,
  Koutra, D., Faloutsos, C., and Li, L. (2012).
\newblock Rolx: structural role extraction \& mining in large graphs.
\newblock In {\em Proceedings of the 18th ACM SIGKDD international conference
  on Knowledge discovery and data mining}, pages 1231--1239. ACM.

\bibitem[Hoff et~al., 2002]{hoff2002latent}
Hoff, P.~D., Raftery, A.~E., and Handcock, M.~S. (2002).
\newblock Latent space approaches to social network analysis.
\newblock {\em Journal of the american Statistical association},
  97(460):1090--1098.

\bibitem[Holland et~al., 1983]{holland1983stochastic}
Holland, P.~W., Laskey, K.~B., and Leinhardt, S. (1983).
\newblock Stochastic blockmodels: First steps.
\newblock {\em Social networks}, 5(2):109--137.

\bibitem[Horn and Johnson, 2012]{horn2012matrix}
Horn, R.~A. and Johnson, C.~R. (2012).
\newblock {\em Matrix analysis}.
\newblock Cambridge university press.

\bibitem[Huang and Gao, 2014]{huang2014clustering}
Huang, Y. and Gao, X. (2014).
\newblock Clustering on heterogeneous networks.
\newblock {\em Wiley Interdisciplinary Reviews: Data Mining and Knowledge
  Discovery}, 4(3):213--233.

\bibitem[Id{\'e} and Kashima, 2004]{ide2004eigenspace}
Id{\'e}, T. and Kashima, H. (2004).
\newblock Eigenspace-based anomaly detection in computer systems.
\newblock In {\em Proceedings of the tenth ACM SIGKDD international conference
  on Knowledge discovery and data mining}, pages 440--449. ACM.

\bibitem[Id{\'e} et~al., 2009]{ide2009proximity}
Id{\'e}, T., Lozano, A.~C., Abe, N., and Liu, Y. (2009).
\newblock Proximity-based anomaly detection using sparse structure learning.
\newblock In {\em SDM}, pages 97--108. SIAM.

\bibitem[Jaru{\v{s}}kov{\'a}, 1997]{jaruvskova1997some}
Jaru{\v{s}}kov{\'a}, D. (1997).
\newblock Some problems with application of change-point detection methods to
  environmental data.
\newblock {\em Environmetrics}, 8(5):469--483.

\bibitem[J{\'e}r{\'e}my et~al., 2009]{jeremy2009sources}
J{\'e}r{\'e}my, T., Catherine, M., Guy, G., and Brynjar, K. (2009).
\newblock Sources of bias in synchronization measures and how to minimize their
  effects on the estimation of synchronicity: application to the uterine
  electromyogram.
\newblock In {\em Recent advances in biomedical engineering}. InTech.

\bibitem[Ji et~al., 2013]{ji2013incremental}
Ji, T., Yang, D., and Gao, J. (2013).
\newblock Incremental local evolutionary outlier detection for dynamic social
  networks.
\newblock In {\em Machine Learning and Knowledge Discovery in Databases}, pages
  1--15. Springer.

\bibitem[Jiang et~al., 2015]{jiang2015link}
Jiang, M., Chen, Y., and Chen, L. (2015).
\newblock Link prediction in networks with nodes attributes by similarity
  propagation.
\newblock {\em arXiv preprint arXiv:1502.04380}.

\bibitem[Khoa and Chawla, 2012]{khoa2012large}
Khoa, N. L.~D. and Chawla, S. (2012).
\newblock Large scale spectral clustering using resistance distance and
  spielman-teng solvers.
\newblock In {\em Discovery Science}, pages 7--21. Springer.

\bibitem[Kleinberg et~al., 1999]{kleinberg1999web}
Kleinberg, J.~M., Kumar, R., Raghavan, P., Rajagopalan, S., and Tomkins, A.~S.
  (1999).
\newblock The web as a graph: measurements, models, and methods.
\newblock In {\em Computing and combinatorics}, pages 1--17. Springer.

\bibitem[Kolaczyk, 2009]{kolaczyk2009statistical}
Kolaczyk, E.~D. (2009).
\newblock {\em Statistical analysis of network data: methods and models}.
\newblock Springer.

\bibitem[Kolda and Bader, 2009]{kolda2009tensor}
Kolda, T.~G. and Bader, B.~W. (2009).
\newblock Tensor decompositions and applications.
\newblock {\em SIAM review}, 51(3):455--500.

\bibitem[Kolda and Sun, 2008]{kolda2008scalable}
Kolda, T.~G. and Sun, J. (2008).
\newblock Scalable tensor decompositions for multi-aspect data mining.
\newblock In {\em Data Mining, 2008. ICDM'08. Eighth IEEE International
  Conference on}, pages 363--372. IEEE.

\bibitem[Koujaku et~al., 2013]{koujaku2013structrual}
Koujaku, S., Kudo, M., Takigawa, I., and Imai, H. (2013).
\newblock Structrual change point detection for evolutional networks.
\newblock In {\em Proceedings of the World Congress on Engineering}, volume~1.

\bibitem[Koujaku et~al., 2015]{koujaku2015community}
Koujaku, S., Kudo, M., Takigawa, I., and Imai, H. (2015).
\newblock Community change detection in dynamic networks in noisy environment.
\newblock In {\em Proceedings of the 24th International Conference on World
  Wide Web Companion}, pages 793--798. International World Wide Web Conferences
  Steering Committee.

\bibitem[Koutra et~al., 2012]{koutra2012tensorsplat}
Koutra, D., Papalexakis, E.~E., and Faloutsos, C. (2012).
\newblock Tensorsplat: Spotting latent anomalies in time.
\newblock In {\em Informatics (PCI), 2012 16th Panhellenic Conference on},
  pages 144--149. IEEE.

\bibitem[Lavielle and Teyssiere, 2006]{lavielle2006detection}
Lavielle, M. and Teyssiere, G. (2006).
\newblock Detection of multiple change-points in multivariate time series.
\newblock {\em Lithuanian Mathematical Journal}, 46(3):287--306.

\bibitem[Lee and Priebe, 2011]{lee2011latent}
Lee, N. and Priebe, C. (2011).
\newblock A latent process model for time series of attributed random graphs.
\newblock {\em Statistical inference for stochastic processes}, 14(3):231--253.

\bibitem[Leskovec et~al., 2005]{leskovec2005realistic}
Leskovec, J., Chakrabarti, D., Kleinberg, J., and Faloutsos, C. (2005).
\newblock Realistic, mathematically tractable graph generation and evolution,
  using kronecker multiplication.
\newblock In {\em Knowledge Discovery in Databases: PKDD 2005}, pages 133--145.
  Springer.

\bibitem[Li et~al., 2016]{li2016social}
Li, Z., Sun, D.-y., Li, J., and Li, Z.-f. (2016).
\newblock Social network change detection using a genetic algorithm based back
  propagation neural network model.
\newblock In {\em Advances in Social Networks Analysis and Mining (ASONAM),
  2016 IEEE/ACM International Conference on}, pages 1386--1387. IEEE.

\bibitem[Lov{\'a}sz, 1993]{asz80random}
Lov{\'a}sz, L. (1993).
\newblock Random walks on graphs: A survey.
\newblock {\em Combinatorics: Paul Erd{\"o}s is}, 80:353--397.

\bibitem[Marchette, 2012]{marchette2012scan}
Marchette, D. (2012).
\newblock Scan statistics on graphs.
\newblock {\em Wiley Interdisciplinary Reviews: Computational Statistics},
  4(5):466--473.

\bibitem[McCulloh and Carley, 2011]{mcculloh2011detecting}
McCulloh, I. and Carley, K.~M. (2011).
\newblock Detecting change in longitudinal social networks.
\newblock Technical report, Military Academy West Point NY Network Science
  Center (NSC).

\bibitem[Meil{\u{a}}, 2003]{meilua2003comparing}
Meil{\u{a}}, M. (2003).
\newblock Comparing clusterings by the variation of information.
\newblock In {\em Learning theory and kernel machines}, pages 173--187.
  Springer.

\bibitem[M{\"u}ller et~al., 2005]{muller2005detection}
M{\"u}ller, M., Baier, G., Galka, A., Stephani, U., and Muhle, H. (2005).
\newblock Detection and characterization of changes of the correlation
  structure in multivariate time series.
\newblock {\em Physical Review E}, 71(4):046116.

\bibitem[Neil et~al., 2013a]{neil2013scan}
Neil, J., Hash, C., Brugh, A., Fisk, M., and Storlie, C.~B. (2013a).
\newblock Scan statistics for the online detection of locally anomalous
  subgraphs.
\newblock {\em Technometrics}, 55(4):403--414.

\bibitem[Neil et~al., 2013b]{neil2013towards}
Neil, J., Uphoff, B., Hash, C., and Storlie, C. (2013b).
\newblock Towards improved detection of attackers in computer networks: New
  edges, fast updating, and host agents.
\newblock In {\em Resilient Control Systems (ISRCS), 2013 6th International
  Symposium on}, pages 218--224. IEEE.

\bibitem[Newman, 2004]{newman2004detecting}
Newman, M.~E. (2004).
\newblock Detecting community structure in networks.
\newblock {\em The European Physical Journal B-Condensed Matter and Complex
  Systems}, 38(2):321--330.

\bibitem[Ng et~al., 2001]{ng2001spectral}
Ng, A.~Y., Jordan, M.~I., and Weiss, Y. (2001).
\newblock On spectral clustering1 analysis and an algorithm.
\newblock {\em Proceedings of Advances in Neural Information Processing
  Systems. Cambridge, MA: MIT Press}, 14:849--856.

\bibitem[Nickel, 2007]{nickel2007random}
Nickel, C. L.~M. (2007).
\newblock {\em Random dot product graphs: A model for social networks},
  volume~68.

\bibitem[Nickel, 2013]{nickel2013tensor}
Nickel, M. (2013).
\newblock {\em Tensor factorization for relational learning}.
\newblock PhD thesis, lmu.

\bibitem[Pandit et~al., 2007]{pandit2007netprobe}
Pandit, S., Chau, D.~H., Wang, S., and Faloutsos, C. (2007).
\newblock Netprobe: a fast and scalable system for fraud detection in online
  auction networks.
\newblock In {\em Proceedings of the 16th international conference on World
  Wide Web}, pages 201--210. ACM.

\bibitem[Pao et~al., 2011]{pao2011statistical}
Pao, H., Coppersmith, G.~A., and Priebe, C.~E. (2011).
\newblock Statistical inference on random graphs: Comparative power analyses
  via monte carlo.
\newblock {\em Journal of Computational and Graphical Statistics},
  20(2):395--416.

\bibitem[Papadimitriou et~al., 2010]{papadimitriou2010web}
Papadimitriou, P., Dasdan, A., and Garcia-Molina, H. (2010).
\newblock Web graph similarity for anomaly detection.
\newblock {\em Journal of Internet Services and Applications}, 1(1):19--30.

\bibitem[Papalexakis et~al., 2012]{papalexakis2012parcube}
Papalexakis, E.~E., Faloutsos, C., and Sidiropoulos, N.~D. (2012).
\newblock Parcube: Sparse parallelizable tensor decompositions.
\newblock In {\em Machine Learning and Knowledge Discovery in Databases}, pages
  521--536. Springer.

\bibitem[Park et~al., 2013]{park2013anomaly}
Park, Y., Priebe, C.~E., and Youssef, A. (2013).
\newblock Anomaly detection in time series of graphs using fusion of graph
  invariants.
\newblock {\em Selected Topics in Signal Processing, IEEE Journal of},
  7(1):67--75.

\bibitem[Peel and Clauset, 2015]{peel2015detecting}
Peel, L. and Clauset, A. (2015).
\newblock Detecting change points in the large-scale structure of evolving
  networks.
\newblock In {\em Twenty-Ninth AAAI Conference on Artificial Intelligence}.

\bibitem[Pincombe, 2005]{pincombe2005anomaly}
Pincombe, B. (2005).
\newblock Anomaly detection in time series of graphs using arma processes.
\newblock {\em ASOR BULLETIN}, 24(4):2.

\bibitem[Priebe et~al., 2005]{priebe2005scan}
Priebe, C.~E., Conroy, J.~M., Marchette, D.~J., and Park, Y. (2005).
\newblock Scan statistics on enron graphs.
\newblock {\em Computational \& Mathematical Organization Theory},
  11(3):229--247.

\bibitem[Ranshous et~al., 2015]{ranshous2015anomaly}
Ranshous, S., Shen, S., Koutra, D., Harenberg, S., Faloutsos, C., and Samatova,
  N.~F. (2015).
\newblock Anomaly detection in dynamic networks: a survey.
\newblock {\em Wiley Interdisciplinary Reviews: Computational Statistics},
  7(3):223--247.

\bibitem[Rossi et~al., 2013]{rossi2013modeling}
Rossi, R.~A., Gallagher, B., Neville, J., and Henderson, K. (2013).
\newblock Modeling dynamic behavior in large evolving graphs.
\newblock In {\em Proceedings of the sixth ACM international conference on Web
  search and data mining}, pages 667--676. ACM.

\bibitem[Salter-Townshend et~al., 2012]{salter2012review}
Salter-Townshend, M., White, A., Gollini, I., and Murphy, T.~B. (2012).
\newblock Review of statistical network analysis: models, algorithms, and
  software.
\newblock {\em Statistical Analysis and Data Mining}, 5(4):243--264.

\bibitem[Shah et~al., 2015]{shah2015timecrunch}
Shah, N., Koutra, D., Zou, T., Gallagher, B., and Faloutsos, C. (2015).
\newblock Timecrunch: Interpretable dynamic graph summarization.
\newblock In {\em Proceedings of the 21th ACM SIGKDD International Conference
  on Knowledge Discovery and Data Mining}, pages 1055--1064. ACM.

\bibitem[Singh and Singh, 2015]{singh2015fraud}
Singh, P. and Singh, M. (2015).
\newblock Fraud detection by monitoring customer behavior and activities.
\newblock {\em International Journal of Computer Applications}, 111(11).

\bibitem[Sricharan and Das, 2014]{sricharan2014localizing}
Sricharan, K. and Das, K. (2014).
\newblock Localizing anomalous changes in time-evolving graphs.
\newblock In {\em Proceedings of the 2014 ACM SIGMOD international conference
  on Management of data}, pages 1347--1358. ACM.

\bibitem[Sun et~al., 2006]{sun2006beyond}
Sun, J., Tao, D., and Faloutsos, C. (2006).
\newblock Beyond streams and graphs: dynamic tensor analysis.
\newblock In {\em Proceedings of the 12th ACM SIGKDD international conference
  on Knowledge discovery and data mining}, pages 374--383. ACM.

\bibitem[Sussman et~al., 2012]{sussman2012consistent}
Sussman, D.~L., Tang, M., Fishkind, D.~E., and Priebe, C.~E. (2012).
\newblock A consistent adjacency spectral embedding for stochastic blockmodel
  graphs.
\newblock {\em Journal of the American Statistical Association},
  107(499):1119--1128.

\bibitem[Tang et~al., 2008]{tang2008community}
Tang, L., Liu, H., Zhang, J., and Nazeri, Z. (2008).
\newblock Community evolution in dynamic multi-mode networks.
\newblock In {\em Proceedings of the 14th ACM SIGKDD international conference
  on Knowledge discovery and data mining}, pages 677--685. ACM.

\bibitem[Tang et~al., 2012]{tang2012community}
Tang, L., Wang, X., and Liu, H. (2012).
\newblock Community detection via heterogeneous interaction analysis.
\newblock {\em Data Mining and Knowledge Discovery}, 25(1):1--33.

\bibitem[Tang et~al., 2013]{tang2013attribute}
Tang, M., Park, Y., Lee, N.~H., and Priebe, C.~E. (2013).
\newblock Attribute fusion in a latent process model for time series of graphs.
\newblock {\em IEEE Transactions on Signal Processing}, 61(7):1721--1732.

\bibitem[Turcotte, 2014]{turcotte2014intended}
Turcotte, M. (2014).
\newblock Intended for: Phd thesis.

\bibitem[Wang et~al., 2008]{wang2008spatial}
Wang, B., Phillips, J.~M., Schreiber, R., Wilkinson, D.~M., Mishra, N., and
  Tarjan, R. (2008).
\newblock Spatial scan statistics for graph clustering.
\newblock In {\em SDM}, pages 727--738. SIAM.

\bibitem[Wang et~al., 2014]{wang2014locality}
Wang, H., Tang, M., Park, Y., and Priebe, C. (2014).
\newblock Locality statistics for anomaly detection in time series of graphs.

\bibitem[Yu et~al., 2014]{yu2014glad}
Yu, R., He, X., and Liu, Y. (2014).
\newblock Glad: Group anomaly detection in social media analysis.
\newblock In {\em Proceedings of the 20th ACM SIGKDD international conference
  on Knowledge discovery and data mining}, pages 372--381. ACM.

\bibitem[Zhao et~al., 2011]{zhao2011community}
Zhao, Y., Levina, E., and Zhu, J. (2011).
\newblock Community extraction for social networks.
\newblock {\em Proceedings of the National Academy of Sciences},
  108(18):7321--7326.

\end{thebibliography}

\end{document}